\documentclass[onecolumn,showpacs,preprintnumbers,amsmath,amssymb,10pt,a4paper]{revtex4}

\usepackage{geometry}
\usepackage{graphicx}% Include figure files
\usepackage{dcolumn}% Align table columns on decimal point
\usepackage{bm}% bold math
\newcommand{\be}{\begin{equation}}
\newcommand{\ee}{\end{equation}}
\newcommand{\bd}{\begin{displaymath}}
\newcommand{\ed}{\end{displaymath}}
\newcommand{\BE}{\begin{eqnarray}}
\newcommand{\EE}{\end{eqnarray}}

\newcommand{\bn}{\ensuremath{\mathbf{n}}}

\newcommand{\avg}[1]{\left\langle{#1}\right\rangle}

\newcommand{\fE}{\mathbb{E}}

\begin{document}

\preprint{}
\title{Intrinsic fluctuations in stochastic delay systems: theoretical description and application to a simple model of gene regulation}
% Force line breaks with \\

\author{Tobias Galla}
\email{tobias.galla@manchester.ac.uk}

\affiliation{Theoretical Physics, School of Physics and Astronomy, The University of Manchester, Manchester M13 9PL, United Kingdom}

\date{\today}% It is always \today, today,

\begin{abstract}
The effects of intrinsic noise on stochastic delay systems is studied
within an expansion in the inverse system size. We show that the
stochastic nature of the underlying dynamics may induce oscillatory
behaviour in parameter ranges where the deterministic system does not
sustain cycles, and compute the power spectra of these stochastic
oscillations analytically, in good agreement with simulations. The
theory is developed in the context of a simple one-dimensional toy
model, but is applicable more generally. Gene regulatory systems in
particular often contain only a small number of molecules, leading to
significant fluctuations in mRNA and protein concentrations. As an
application we therefore study a minimalistic model of the expression
levels of $hes1$ mRNA and Hes1 protein, representing the simple motif
of an auto-inhibitory feedback loop and motivated by its relevance to
somite segmentation.
\end{abstract}

\maketitle
%%%%%%%%%%%%%%%%%%%%%%%%%%%%%%%%%%%%%%%%%%%%%%%%%%%%%%%%%%%%%%%%%%%%%%%%%%%%%%%%
%%%%%%%%%%%%%%%%%%%%%%%%%%%%%%%%%%%%%%%%%%%%%%%%%%%%%%%%%%%%%%%%%%%%%%%%%%%%%%%%
\section{Introduction}
Most processes in biology are intrinsically stochastic, due to the
random fashion in which molecules interact. In order for a biochemical
reaction to occur, for example, all reagents must be sufficiently
close in space, and due to thermal or other types of stochasticity the
execution of reaction is fundamentally a stochastic process. This type
of randomness has, until recently, mostly been neglected in attempts
to model biochemical reaction systems, and deterministic rate
equations have often been used to describe the dynamics of such
systems. Noise has here often been assumed to have only a minor effect on
the dynamics, so that it could safely be ignored. The use of
deterministic approaches implies the assumption of large, formally
infinite system sizes, only in this limit can the law of large numbers
be applied to show that the resulting mean-field equations give an
accurate description of the dynamics of the system. Additionally, it
is frequently assumed implicitly that the reactor in which the chemical dynamics
takes place is well-mixed, so that spatial variation of concentrations
of the interacting chemicals can be ignored.

The reason for the popularity of such approaches undoubtedly rests in
their relative mathematical simplicity: while the methods with which
to analyse sets of non-linear deterministic differential equation are
fully developed (see e.g. \cite{strogatz} or similar textbooks), a
theory for the corresponding stochastic systems is far less
advanced. If the number of reacting molecules in the system is small,
the stochastic effects can no longer be ignored. An important example
are mRNA molecules in gene expression \cite{kaern,bratsun, barrio,
schlicht}, where only a small number of molecules is involved in the
reaction dynamics. Unsurprisingly deviations from the mean-field
dynamics are here to be expected, and stochastic rather than
deterministic modeling approaches to such systems in molecular
biology are appropriate.  Only in recent years have analytical and
more systematic studies of such systems been undertaken, and
substantial differences between the behaviour of stochastic systems
and their deterministic counterparts have been found in different
model systems. In particular so-called demographic noise \cite{nisbet}
due to the discreteness of the dynamics may change the structure of
the attractor of a given system fundamentally. References relevant for
the present work can be found e.g. in \cite{schlicht, alan1, alan2,
barrio, bratsun, various}.

Stochastic approaches to biochemical reaction systems typically start
from a master equation describing the microscopic dynamics, the
deterministic mean-field dynamics can then formally be derived to
lowest order within a van Kampen expansion in the inverse system
size \cite{vankampen}. Taking into account next-to-leading order finite-size
corrections can alter the dynamics considerably. Predator-prey systems with a fixed point on the deterministic level can for
example be seen to exhibit coherent sustained oscillations at finite
sizes \cite{alan1}. The spectrum of these cycles can be obtained to
striking precision within the system-size expansion. Similar
oscillations have been found in a variety of other systems, including
models of epidemiology, opinion dynamics and biochemical reaction
networks \cite{alan2,various}. 

In addition to the discreteness of the dynamics and the resulting
intrinsic stochasticity, processes in gene regulatory systems are
typically subject to considerable delays induced by the underlying
biochemical reactions. I.e. processes such as transcription and
translation do not occur instantaneously, but generate their reaction
products only well after the reaction has been triggered
\cite{bratsun, barrio, schlicht}. The aim of the present paper is
therefore to extend the theoretical tools developed in
\cite{alan1,alan2,various} to the case of stochastic delay systems, and to
use them to study a simple model of gene regulation. As we will see
the dynamics of such systems may well exhibit stochastic coherent
oscillations at finite system size in ranges of the reaction rates
where the deterministic, infinite delay system has no cycles, but approaches
a fixed point instead. Such oscillations in delay stochastic systems
have been reported in \cite{bratsun, barrio, schlicht}, but to our
knowledge a theoretical computation of correlations and power spectra of these cycles within a systematic expansion in the inverse system size has not
been attempted in the context of such models. Theoretical approaches
based on generating functions have however been discussed in
\cite{bratsun}. In this paper we will follow a complementary approach,
and in particular we will describe how the method of the system-size
expansion applies to delay systems, and how a linear delay Langevin equation
can be derived for fluctuations about the trajectory of the
deterministic mean-field system. From this Langevin equation the power
spectra of these stochastic cycles can then be obtained analytically. Our analysis therefore offers a theoretical characterization of results from simulations reported e.g. in \cite{barrio}, and an alternative to the analytical approaches discussed in \cite{bratsun}.
\section{Toy model}
\subsection{Definition and deterministic description}
In order to develop the formalism we start with a simple model of delay stochastic processes, and consider a system in which there is only one reacting substance $X$. The dynamics are given by the following reactions
\BE
A&\longrightarrow&A+X,\label{eq:1}\\
B+X&\longrightarrow&B,\label{eq:2}\\
X+C&\Longrightarrow&C,\label{eq:3}
\EE
note that neither reaction affects the number molecules of substances $A,B$ and $C$ in the system, so that these reactants are mere `dummy' variables, and their only role is to set the relative rates with which the three reactions occur. While the first two reactions are assumed to occur instantaneously, the third reaction involves a delay, we indicate this by the double arrow in Eq. (\ref{eq:3}). I.e. if a reaction between a molecule of type $C$ and a molecule of type $X$ is triggered at time $t$, then one molecule of type $X$ is removed from the system at a later time $t+\tau$ (provided there is at least one $X$-molecule in the system at this later time). For simplicity we will assume that $\tau$ is a constant delay period, but variable delay times, drawn e.g. from some probability distribution can be in principle considered as well \cite{schlicht}. The model in this setup has previously been discussed and studied within an alternative approach in \cite{bratsun}.

On the mean-field level the concentration of $X$-molecules in the system is described by the delay differential equation
\be
\dot x(t)=a-bx(t)-cx(t-\tau),\label{eq:simple}
\ee
where $a,b$ and $c$ are non-negative rate constants related to the (constant) number of molecules of types $A,B$ and $C$ in the system respectively (further details will be discussed below).

Eq. (\ref{eq:simple}) is linear and its asymptotic behaviour can be
computed straightforwardly. In particular a linear stability analysis
has been carried out in \cite{bratsun}, and a phase diagram was
obtained in terms of the parameters $a,b,c$ and $\tau$. The only
fixed point of Eq. (\ref{eq:simple}) is $x^*=a/(b+c)$, and at fixed $a$ it
is found to be unstable at large $\tau c$ or small $\tau b$
respectively. In such circumstances oscillations grow
indefinitely. Below a the line marking the Hopf bifurcation in the $(\tau
b, \tau c)$ plane, the fixed-point is stable \cite{bratsun}. We will focus on this
regime in the following.

\subsection{Stochastic dynamics, van Kampen expansion and spectrum of fluctuations}
In order to model the dynamics on the microscopic level, let us assume the reactor in which the various reactions take place contains $a\Omega$ molecules of type $A$, $b\Omega$ molecules of type $B$, $c\Omega$ molecules of type $C$, and $n$ particles of type $X$. Since the number of $A,B$ and $C$ molecules remains unchanged by the reactions given above, the only dynamical variable in the system is $n(t)$ (which will be of order $\Omega$ as are the numbers of the other particles in the system). The first reaction, Eq. (\ref{eq:1}), then occurs with a rate $a\Omega$, the second reaction with rate $nb$ and the third with rate $nc$, and the resulting stochastic process is described by the master equation \cite{bratsun}
\BE\label{eq:master}
\frac{d}{dt} P(n,t)&=&a\Omega(\fE^{-1}-1) P(n,t)+b (\fE-1) [n P(n,t)]\nonumber \\
&&+c \sum_{m=0}^\infty m (\fE-1)[\Theta(n) P(n,t;m,t-\tau)]
\EE
for the probability $P(n,t)$ of finding the system in state $n$ at time $t$. $\fE$ is here an operator acting on a function of $n$ via $\fE f(n)=f(n+1)$, and not to be confused with an expectation value of some kind. $\fE^{-1}$ stands for the inverse operation. $\Theta(n)$ is the step function, i.e. $\Theta(n>0)=1$ and $\Theta(n=0)=0$ which ensures that the delayed removal of $X$-molecules only occurs provided there is at least one molecule present in the system at the time at which the removal is due to take place. Note that Eq. (\ref{eq:master}) is not closed on the level of one-time quantities, as $P(n,t;m,t-\tau)$ describes the joint probability distribution of finding $n$ $X$-molecules at time $t$, and $m$ $X$-molecules at time $t-\tau$.

Following \cite{vankampen} and anticipating that $n$ will be of order $\Omega$ with fluctuations of order $\Omega^{1/2}$ we now introduce continuous degrees of freedom, and write
\be
\frac{n(t)}{\Omega}=x(t)+\frac{\xi(t)}{\Omega^{1/2}}.
\ee
The above master equation for $P(n,t)$ can then be written in terms of the distribution $\Pi(\xi,t)$, and within an expansion in powers of $\Omega^{-1/2}$ we have similar to \cite{vankampen}
\BE\label{eq:expand}
&&\partial_t \Pi(\xi,t)-\Omega^{1/2}\frac{\partial\Pi(\xi,t)}{\partial\xi}\dot x\nonumber \\
&=&a\Omega\left[-\Omega^{-1/2}\frac{\partial}{\partial \xi}+\Omega^{-1}\frac{1}{2}\frac{\partial^2}{\partial \xi^2}\right] \Pi(\xi,t)+b\left[\Omega^{-1/2}\frac{\partial}{\partial \xi}+\Omega^{-1}\frac{1}{2}\frac{\partial^2}{\partial \xi^2}\right] \left[(\Omega x(t)+\Omega^{1/2}\xi)\Pi(\xi,t)\right]\nonumber \\
&&+c\int d\eta \left\{\left[\Omega^{-1/2}\frac{\partial}{\partial \xi}+\Omega^{-1}\frac{1}{2}\frac{\partial^2}{\partial \xi^2}\right]\left[(\Omega x(t-\tau)+\Omega^{1/2}\eta)\Pi(\xi,t;\eta,t-\tau)\right]\right\},
\EE
where we introduce $\eta$ by writing  $n(t-\tau)/\Omega=x(t-\tau)+\eta/\Omega^{1/2}$, and where we have ignored higher-order terms. Anticipating that we will take the limit of large systems eventually and that trajectories at which $n(t)=0$ at any time will not contribute in this limit we have ignored the factor $\Theta(n)$ in the last term of (\ref{eq:master}). This is common in the context of the van Kampen expansion, which is usually unable to capture features such as absorbing states at the boundaries of configuration space. Collecting terms of order $\Omega^{1/2}$ in Eq. (\ref{eq:expand}) one finds
\BE\label{eq:lowest}
-\Omega^{1/2}\frac{\partial\Pi}{\partial\xi}\dot x&=&-a\Omega^{1/2}\frac{\partial\Pi}{\partial\xi}+bx(t)\Omega^{1/2}\frac{\partial\Pi}{\partial\xi}+cx(t-\tau)\Omega^{1/2}\frac{\partial\Pi}{\partial\xi}\EE
in the lowest order of the van Kampen expansion. Wh have here written $\Pi$ as a shorthand for $\Pi(\xi,t)$ and used the identity $\Pi(\xi,t)=\int d\eta~ \Pi(\xi,t;\eta,t-\tau)$. From (\ref{eq:lowest}) one has
\be
\dot x(t)=a-bx(t)-cx(t-\tau),\label{eq:mf}
\ee
i.e. one recovers the above deterministic equation (\ref{eq:simple}). To next-leading order (i.e. collecting ${\cal O}(\Omega^0)$ terms) one finds
\BE
\partial_t \Pi(\xi,t)&=&\frac{1}{2}a\frac{\partial^2}{\partial \xi^2}\Pi(\xi,t)+b\frac{\partial}{\partial\xi}\left[\xi \Pi(\xi,t)\right]+\frac{1}{2}bx(t)\frac{\partial^2}{\partial \xi^2}\Pi(\xi,t)\nonumber \\
&&+c\int d\eta\left\{\frac{\partial}{\partial\xi}\left[\eta \Pi(\xi,t;\eta,t-\tau)\right]+\frac{1}{2}x(t-\tau)\frac{\partial^2}{\partial \xi^2}\Pi(\xi,t;\eta,t-\tau)\right\}.
\EE
Integrating out $\eta$ in the last term one then has
\BE\label{eq:expand2}
\partial_t \Pi(\xi,t)&=&\frac{1}{2}a\frac{\partial^2}{\partial \xi^2}\Pi(\xi,t)+b\frac{\partial}{\partial\xi}\left[\xi \Pi(\xi,t)\right]+\frac{1}{2}bx(t)\frac{\partial^2}{\partial \xi^2}\Pi(\xi,t)\nonumber \\
&&+c\int d\eta\left\{\frac{\partial}{\partial\xi}\left[\eta \Pi(\xi,t;\eta,t-\tau)\right]\right\}+c\frac{1}{2}x(t-\tau)\frac{\partial^2}{\partial \xi^2}\Pi(\xi,t).
\EE
At asymptotic times $t$ the mean-field trajectory approaches its fixed point (our analysis is restricted to the stable phase, for similar studies in non-delay systems with a limit cycle see \cite{boland}). We therefore replace $x(t)$ and $x(t-\tau)$ in Eq. (\ref{eq:expand2}) by $x^*=a/(b+c)$. With this substitution Eq. (\ref{eq:expand2}) then describes a delayed Langevin dynamics of the form
\be\label{eq:langevin}
\dot \xi = - b \xi(t) - c \xi(t-\tau)+ \zeta(t),
\ee
where $\zeta(t)$ is Gaussian white noise of zero mean and with variance
\be
\avg{\zeta(t)\zeta(t')}=(a+bx^*+cx^*)\delta(t-t').\ee
See e.g. Frank et al. \cite{frank} for further details on Fokker-Planck descriptions of delay systems. Eq. (\ref{eq:langevin}) is linear and can be solved in Fourier space (similar approaches to delay Langevin equations have been discussed in \cite{guillouzic}). In particular one has
\be\label{eq:fourier}
[i\omega+b+ce^{-i\omega\tau}]\widetilde\xi(\omega)=\widetilde\zeta(\omega),
\ee
where $\widetilde\xi(\omega)$ and $\widetilde\zeta(\omega)$ indicate the Fourier transforms of $\xi(t)$ and $\zeta(t)$ respectively. From Eq. (\ref{eq:fourier}) one directly reads off the power spectrum of $\xi(t)$ and finds
\BE\label{eq:spec}
P(\omega)&\equiv&\avg{|\widetilde\xi(\omega)|^2}\nonumber\\
&=&\frac{a+bx^*+cx^*}{[b+c\cos(\omega\tau)]^2+[\omega-c\sin(\omega\tau)]^2}=\frac{2a}{[b+c\cos(\omega\tau)]^2+[\omega-c\sin(\omega\tau)]^2}.
\EE
We have here used the relation $\avg{\widetilde\zeta(\omega)\widetilde\zeta(\omega')}=(a+bx^*+cx^*)\delta(\omega+\omega')$, where $\avg{\cdots}$ denotes an average over the stochastic process described by the Fokker-Planck equation (\ref{eq:expand2}), or equivalently over realizations of the Langevin equation (\ref{eq:langevin}).
 \subsection{Test against simulations}
\begin{figure}[t]
\centerline{\includegraphics[width=0.5\textwidth]{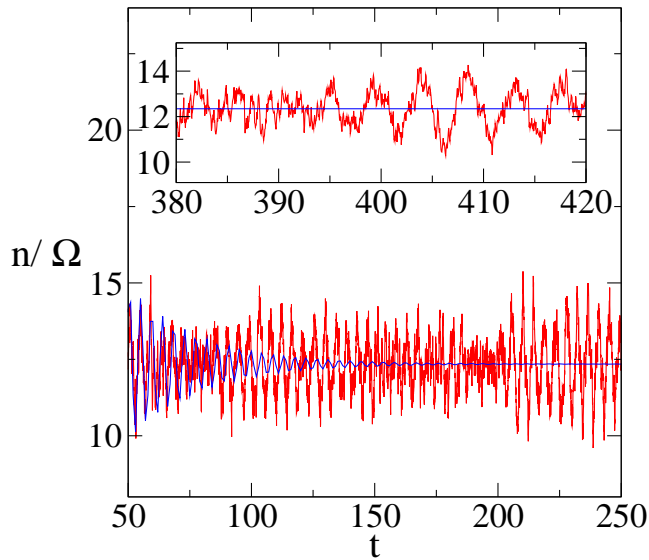}}
\vspace{1em}
\caption{(Colour on-line). Dynamics of the toy model at $a=100, b=4.1, c=4., \tau=2$. The dark line with decaying oscillations in the main panel shows the behaviour of the concentration of $X$ molecules in the deterministic system, Eq. (\ref{eq:mf}) \cite{discl}. The noisy line with persistent oscillations represents one simulation run of the stochastic dynamics at $\Omega=100$. The inset shows a zoom at large times in the equilibrated regime, the horizontal line is the mean-field fixed point, the stochastic system shows persistent cycles.}
\label{fig:fig1}
\end{figure}
Microscopic simulations of the processes defined by the reactions (\ref{eq:1}-\ref{eq:3}) can be carried out using the algorithm originally proposed by Gillespie \cite{gillespie}, suitably modified to take into account the delayed reactions. Details of such modified Gillespie schemes have been discussed for example in \cite{anderson}, but for completeness we re-iterate them here. Essentially the simulations follow that of the classic Gillespie algorithm \cite{gillespie}, and whenever a delayed reaction is triggered it is added to a list of delay reactions to be executed at a later time ($\tau$ units of time after the reaction is initiated). This list is constantly updated, and delay reactions are executed (and removed from the list) in a manner consistent with the probabilistic description in terms of the above master equation. Specifically the simulations of our toy model dynamics proceed according to the following algorithm:
\begin{enumerate}
\item[1.] Initialise. Set model parameters $a,b,c$ and the system size $\Omega$. Set the initial number $n$ of molecules $X$, and set $t=0$. Set list of scheduled delay reaction to an empty list.
\item[2.] Calculate the propensity functions $a_1=a\Omega$, $a_2=b n$, $a_3=c n$.
\item[3.] Compute $a_0=a_1+a_2+a_3$.
\item[4.] Generate an independent random number $r$ from a uniform distribution over $(0,1]$, and set $\Delta=-\ln(r)/a_0$.
\item[5.] If there is a delayed reaction scheduled to occur during the interval $[t,t+\Delta)$ then
\begin{enumerate}
\item Identify next delayed reaction scheduled, and, provided $n>0$ execute it, i.e. reduce $n$ by one. If $n=0$ before the reaction, then do not execute the update (otherwise $n$ would go negative). In either case remove the reaction from the list of scheduled reactions.
\item Update $t$ to the time for which this reaction was scheduled.
\item Go to 2.
\end{enumerate}
\item[6.] If there is no delayed reaction scheduled for the interval $[t,t+\Delta)$ then 
\begin{enumerate}
\item Generate an independent random number $r$ from a uniform distribution over $(0,1]$, and find $\mu\in\{1,2,3\}$ such that
\begin{displaymath}
\sum_{k=1}^{\mu-1}a_k<r'\leq\sum_{k=1}^{\mu}a_k.
\end{displaymath}
\item If $\mu=1$ or $\mu=2$ and then execute the corresponding reaction (not a delay reaction), and increment $t$ by $\Delta$. Go to 2.  
\item If $\mu=3$, schedule a reaction of type $X+C\Longrightarrow C$ to be executed at later time $t+\tau$, i.e. amend list of scheduled reactions accordingly. Increment $t$ by $\Delta$, and go to 2.
\end{enumerate}
\end{enumerate}
Each run of the simulation generates a time series $n(t)$ from which $\xi(t)=\sqrt{N}[n(t)/\Omega-x^*]$ can be obtained (after a suitable equilibration time), where $x^*$ is the asymptotic fixed-point value of the deterministic dynamics given by Eq. (\ref{eq:mf}), i.e. $x^*=a/(b+c)$, or equivalently the long-time average of $n(t)/\Omega$. From these time series $\xi(t)$ a numerical measurement of the power spectrum $P(\omega)$ is obtained by subsequent Fourier transform, and finally results are averaged over a sufficiently large number of independent runs.

\begin{figure}[t]
\vspace{3em}
\centerline{\includegraphics[width=0.4\textwidth]{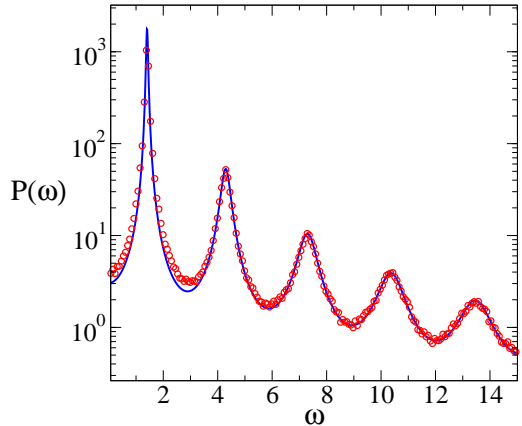}}
\vspace{0.5em}
\caption{(Colour on-line) Power spectrum $P(\omega)=\avg{|\widetilde\xi(\omega)|^2}$ of the fluctuations about the mean-field fixed point of the toy model. Parameters $a,b$ and $c$ are as in Fig. \ref{fig:fig1}. The line shows the analytically obtained spectrum of (\ref{eq:spec}), symbols represent results from simulations at $\Omega=200$ (averaged over $341$ samples, measurements start at $t=100$ to allow for some equilibration period).}
\label{fig:fig2}
\end{figure}
Results of stochastic simulations of this system are shown in
Figs. \ref{fig:fig1} and \ref{fig:fig2}. The first figure depicts a
single run of the stochastic system and shows that coherent
oscillations are sustained in parameter regimes in which the
deterministic equations approach a fixed point. The mechanism by which
these oscillations are generated is the following: the deterministic
system approaches its fixed point in an oscillatory manner (i.e. the
Jacobian at the fixed point has eigenvalues with non-zero imaginary
parts), and the stochasticity of the finite system results in
persistent perturbations away from this fixed point, so that both
features together result in an overall oscillatory effect. This has
been seen in a variety of different systems \cite{alan1,various}, but
it is worth pointing out that in non-delay systems a minimum of two
dimensions is necessary to allow for a complex eigenvalue of the
Jacobian. In delay systems one degree of freedom is sufficient
\cite{bratsun}, so that even the one-dimensional toy model discussed
in this section is able to produce demographic oscillations about the
mean-field fixed point. Fig. \ref{fig:fig2} demonstrates that the
analytically obtained spectrum, Eq. (\ref{eq:spec}), agrees very well with simulations, we attribute the remaining small discrepancies to finite-size or equilibration effects. A similar figure was obtained by different methods (based on generating functions) in \cite{bratsun}.

%%%%%%%%%%%%%%%%%%%%%%%%%%%%%%%%%%%%%%%

\section{Simple model of gene regulation}
The second system we will be studying is a simple model of gene
regulation. We here chose a system that represents one of the most
common motifs in gene regulatory networks, namely a model of a single
gene-protein synthesis with negative delayed feedback
\cite{bratsun}. Specifically we address a model previously discussed in
\cite{barrio}, describing the coupled time behaviour of the expression
levels of so-called $hes1$ messenger RNA (mRNA) and Hes1
protein. Following the notation in the biology literature we will
italicize and use lower case when referring to mRNA, and will use
non-italicized font with the first letter in upper case when referring
to the protein \cite{hirata, barrio}. Hes1 here is a Notch-signalling
molecule, where the so-called Delta-Notch signalling process is a
mechanism for cell-cell communication and underlies cell
differentiation for example in vertebrates \cite{jian, saga,
schlicht}. Cycles with a time-period of approximately two hours has
been reported for the concentrations of $hes1$ mRNA and Hes1 protein
in mice \cite{hirata, barrio}. These oscillatory processes, also
referred to as the somite-segmentation clock \cite{saga}, are linked
to the formation of somites, i.e. the emergence of blocks of cells
which determine the future positions of skeletal muscles or vertebrae
\cite{schlicht,saga}. Spatial segmentation in the body here can be
understood as a reflexion of temporal oscillations in gene expression
\cite{saga,schlicht,jian,lewis}. The underlying molecular mechanism
producing the oscillations of mRNA and protein are hence of great
interest, and several mathematical models have been proposed, 
among them \cite{lewis,monk,barrio,bratsun} and \cite{jensen}. See also \cite{kaern,scott} for stochastic effects in models of gene regulation.

\begin{figure}[t]
\centerline{\includegraphics[width=0.5\textwidth]{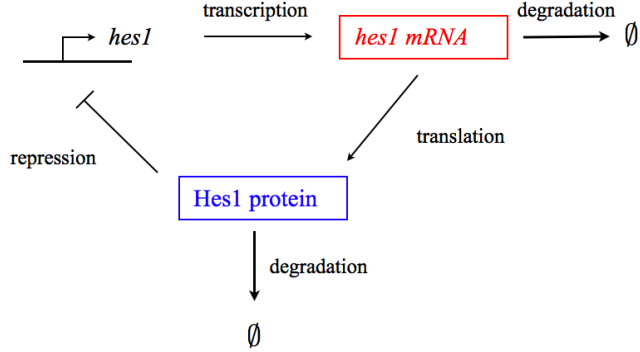}}
\vspace{2em}
\caption{(Colour on-line) Schematic illustration of the Hes1 regulatory system. See \cite{monk} for a similar picture. The first process (transcription)  includes the elongation, splicing, processing and export from the nucleus of primary gene transcript. The synthesis of Hes 1 protein occurs by translation of $hes 1$ mRNA. Any translational delay is here absorbed into the transcriptional delay time $\tau$. The Hes 1 protein finally represses the transcription through the binding to the promoter. Both, the mRNA and the protein are subject to degradation.}
\label{fig:gene}
\end{figure}

We will not discuss the details of the biochemical mechanisms in this
paper, but will only present a brief abstraction of the reactions
necessary to define the mathematical model we will study here. Further
details on modelling genetic circuits can be found in \cite{alon} or
in similar textbooks. In essence the model describes the
concentrations and interactions of two types of substances, $hes1$
mRNA and Hes1 protein, as illustrated in Fig. \ref{fig:gene}. mRNA
molecules are produced by transcription of DNA. This involves several
biochemical processes (e.g. elongation, splicing) which we will
neglect in our description. The rate at which mRNA molecules are
produced depends on the concentration of protein through a negative
feedback mechanism. Transcription is here associated with a delay time
$\tau$, so that it is the protein concentration at time $t-\tau$ which
affects the production rate of mRNA at time $t$. This will be
explained further below. mRNA is also subject to degradation
(i.e. removal from the system) at a constant rate $\mu_m$. In a
process subsequent to transcription $hes1$ mRNA molecules are
translated into Hes1 protein (the mRNA molecule is not used up in this
process). Translation may involve another delay, which for simplicity
can be absorbed into the transcriptional delay \cite{bratsun}. Protein
molecules finally are subject to a degradation process at rate
$\mu_p$. Crucially, a negative feedback is induced by a repressatory
effect of Hes1 protein on the transcription process. Hes1 dimers may
bind to the relevant promoter regions in the DNA, and reduce the
transcription rate at which mRNA is generated (following \cite{barrio}
dimerization is not discussed as an explicit step in our work, but see
\cite{bratsun} for models taking this into account). The
transcriptional repressor Hes1 thus negatively affects its own
expression \cite{hirata}. Mathematically this is modelled by a
transcription rate which depends on the concentration of protein via a
decreasing function, as we will now explain.
We will focus on the model proposed in \cite{monk,barrio}. In its deterministic form it is given by the differential equations
\BE
\frac{d}{dt} M(t)&=&\alpha_m f(P(t-\tau))-\mu_m M(t), \label{eq:barrio1}\\
\frac{d}{dt} P(t)&=&\alpha_p M(t)-\mu_p P(t) \label{eq:barrio2}.
\EE

$M(t)$ here labels the concentration of $hes1$ mRNA, and $P(t)$ that
of the Hes 1 protein. $\mu_m$ and $\mu_p$ are degradation rates for
the mRNA and the protein respectively, and $\alpha_m$ is the mRNA transcription rate in the absence of protein
expression ($f(P)$ is still to be defined, but we will have $f(0)=1$). $\alpha_p$ is the translation rate. $f(P)$
finally is a monotonically decreasing Hill function representing the
suppression of mRNA production through the binding of Hes 1 protein
dimers into the promotion region. In the model it takes the form
\cite{barrio,alon}

\be
f(P(t))=\frac{1}{1+[P(t)/P_0]^h}
\ee
with $h$ the so-called Hill coefficient. $P_0$ is the concentration of protein at which $f(P=P_0)=1/2$. Eqs. (\ref{eq:barrio1},\ref{eq:barrio2}) are the deterministic abstraction of an underlying microscopic stochastic model defined by the following four reactions \cite{barrio}
\BE
M&\stackrel{\mu_m}{\longrightarrow}&\emptyset, \label{eq:r1}\\
P&\stackrel{\mu_p}{\longrightarrow}&\emptyset, \label{eq:r2}\\
M&\stackrel{\alpha_p}{\longrightarrow}&M+P, \label{eq:r3} \\
\emptyset&\stackrel{\alpha_m f}{\Longrightarrow}&M.\label{eq:r4}
\EE

\begin{figure}[t]
\centerline{\includegraphics[width=0.5\textwidth]{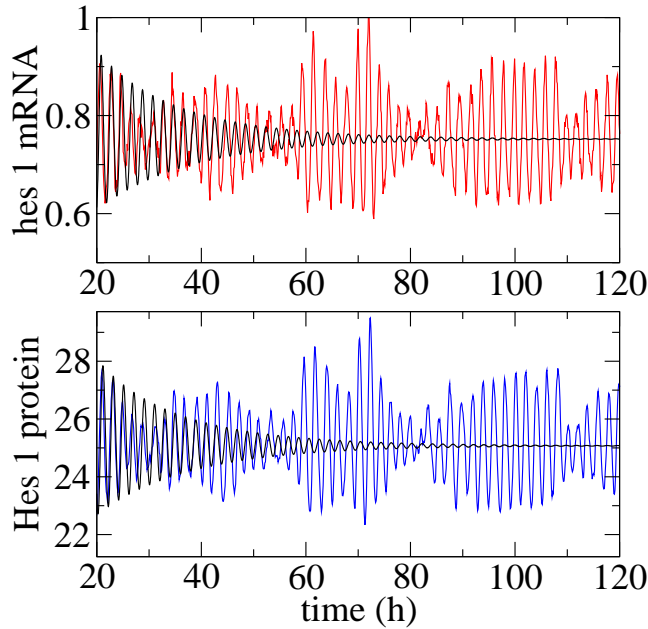}}
\caption{Time series of the concentrations of mRNA and protein concentrations respectively. Solid curves, decaying towards a fixed point, are from a numerical integration of the deterministic dynamics, Eqs. (\ref{eq:barrio1},\ref{eq:barrio2}). Curves with persistent oscillations represent a single run of the stochastic dynamics at $\Omega=1000$. Model parameters are as some of the examples in \cite{barrio}: $P_0=10, h=4.1, \tau=18.7, \alpha_m=\alpha_p=1, \mu_m=\mu_p=0.03$. Units of $\alpha_p,\alpha_m$ and of $\mu_m,\mu_p$ are $min^{-1}$, $\tau$ is measured in minutes \cite{barrio}. }
\label{fig:fig3}
\end{figure}

These dynamics may be described by a stochastic process for the
numbers $n_m$ of mRNA molecules and $n_p$ of protein molecules in the
system. For later convenience we will write $\bn=(n_m,n_p)$. The first
two reactions here describe the degradation of mRNA and protein
respectively. The third reaction captures the translation of mRNA into
protein. The last reaction finally corresponds to the production of
$hes1$ mRNA via transcription. Note that DNA is not part of the
dynamical model (its concentration is constant), which is why $M$
appears to be produced out of the void in the fourth reaction. In
absence of protein ($n_p=0$) this reaction occurs at a rate
$\alpha_m$, but is suppressed by the Hes1 protein, and in total the
rate at which mRNA is produced at time $t$ is hence $\alpha_m
f(n_p(t-\tau)/\Omega)$, where $\Omega$ is a measure of the system
size. The time lag $\tau$ in the argument of $f$ models the delayed
repression of $hes1$ mRNA production by the protein. The rate of
production of mRNA at time $t$ is therefore negatively regulated by
the concentration of protein at time $t-\tau$. A typical run of the stochastic system is shown and compared with the deterministic system in Fig. \ref{fig:fig3}. Model parameters are chosen as in \cite{barrio}. As seen in the figure, the deterministic system approaches a stable fixed point asymptotically, with a complex eigenvalue as indicated by the decaying modulations. The stochastic system at finite size remains in an oscillatory state as previously observed in \cite{barrio}. We will now proceed to characterize these oscillations analytically, applying the formalism developed in the previous section.

The master equation describing the processes (\ref{eq:r1}-\ref{eq:r4}) then takes the form
\BE\label{eq:mastergene}
\frac{d}{dt} P(\bn;t)&=&\mu_m (\fE_M-1)[n_mP(\bn;t)]+\mu_p (\fE_P-1)[n_pP(\bn;t)]+\alpha_p(\fE_P^{-1}-1)[n_mP(\bn;t)] \nonumber \\
&&+\alpha_m\Omega\sum_{\bn'}f(n_p'/\Omega)(\fE_M^{-1}-1)[P(\bn,t;\bn',t-\tau)],
\EE
where $P(\bn,t)$ is the probability of finding the system in state $\bn$ at time $t$, and $P(\bn,t;\bn',t')$ is the probability for the system to be in state $\bn$ at $t$ and in state $\bn'$ at time $t'$. $\fE_M$ and $\fE_P$ are raising operators acting on functions of $n_m,n_p$ via $\fE_M g(n_m,n_p)=g(n_m+1,n_p)$ and  $\fE_P g(n_m,n_p)=g(n_m,n_p+1)$. In the case of two-time quantities, e.g. $P(\bn,t;\bn',t')$ the raising applies with respect to the second argument $\bn'$. Note that all terms on the right-hand side of Eq. (\ref{eq:mastergene}) are of order $\Omega$. This overall factor could in principle be absorbed into a re-scaling of time, even though we will not do so here.

The further analysis proceeds along the lines of what was discussed for the toy model, and we will not report all intermediate steps in all detail. First one writes 
\BE
\frac{n_m(t)}{\Omega}&=&M(t)+\frac{\xi_m(t)}{\Omega^{1/2}},\\
\frac{n_p(t)}{\Omega}&=&P(t)+\frac{\xi_p(t)}{\Omega^{1/2}},
\EE
and then systematically expands the above master equation in powers of $\Omega^{-1/2}$. To lowest order one recovers the mean-field equations (\ref{eq:barrio1},\ref{eq:barrio2}), and in next-to-leading order one finds Langevin equations for the fluctuations about the mean field trajectory. In the fixed-point regime of the mean-field dynamics (i.e. at large times $t$) these equations read 
\BE
\dot \xi_m(t) &=& -\mu_m \xi_m(t)+\alpha_m f'(P^*)\xi_p(t-\tau)+\zeta_m(t) \\
\dot \xi_p(t) &=& \alpha_p \xi_m(t)-\mu_p\xi_p(t)+\zeta_p(t),
\EE
where $(M^*,P^*)$ is the mean-field fixed point, $f'(P)=df(P)/dP=-hP_0^{-1}(1+P/P_0)^{-(h+1)}$. $\zeta_m(t)$ and $\zeta_p(t)$ are Gaussian noise terms of zero mean and in the limit of large $t,t'$ (when the mean-field trajectory has reached its fixed point) they have covariances
\BE
\avg{\zeta_m(t)\zeta_m(t')}&=&\delta(t-t')[\mu_m M^*+\alpha_m f(P^*)], \\
\avg{\zeta_p(t)\zeta_p(t')}&=&\delta(t-t')[\mu_p P^*+\alpha_p M^*],\\
\avg{\zeta_m(t)\zeta_m(t')}&=&0.
\EE
Inverting in Fourier space one then finds after some algebraic manipulations
\BE
\avg{|\widetilde\xi_m(\omega)|^2}&=&\frac{(\omega^2+\mu_p^2)(\alpha_m f(P^*)+\mu_m M^*)+(\alpha_m f'(P^*))^2 (\alpha_p M^*+\mu_p P^*)}{(-\omega^2+\mu_m\mu_p-\alpha_m\alpha_p f'(P^*)\cos(\omega\tau))^2+((\mu_m+\mu_p)\omega+\alpha_m\alpha_p f'(P^*)\sin(\omega\tau))^2},\label{eq:spec1}\\
\avg{|\widetilde\xi_p(\omega)|^2}&=&\frac{\alpha_p^2(\alpha_m f(P^*)+\mu_m M^*)+(\omega^2+\mu_m^2)(\alpha_p M^*+\mu_p P^*)}{(-\omega^2+\mu_m\mu_p-\alpha_m\alpha_p f'(P^*)\cos(\omega\tau))^2+((\mu_m+\mu_p)\omega+\alpha_m\alpha_p f'(P^*)\sin(\omega\tau))^2}.\label{eq:spec2}
\EE 
\begin{figure}[t]
\centerline{\includegraphics[width=0.5\textwidth]{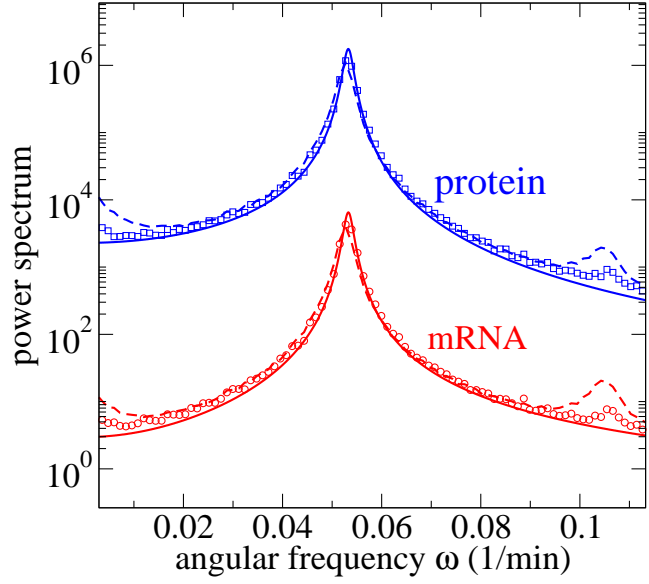}}
\caption{(Colour on-line) Power spectra of the fluctuations of mRNA and protein concentrations respectively. Solid lines are the analytical expressions of Eqs. (\ref{eq:spec1},\ref{eq:spec2}). Markers are from simulations at system size $\Omega=5000$, and dashed lines from simulations at $\Omega=500$. Averages over more than $700$ independent samples are taken in the simulations. Measurements start at $t=3000$min in order to allow for equilibration. Model parameters are $P_0=10, h=4.1, \tau=18.7, \alpha_m=\alpha_p=1, \mu_m=\mu_p=0.03$ \cite{barrio}.}
\label{fig:fig4}
\end{figure}

\begin{figure}[t]
\centerline{\includegraphics[width=0.4\textwidth]{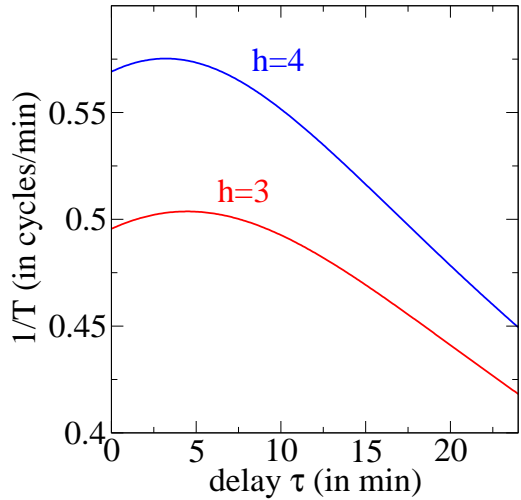}}
\caption{(Colour on-line) Frequency $1/T$ at which the power spectrum of mRNA fluctuations has its maximum. Results are from Eq. (\ref{eq:spec1}), evaluated at the fixed point of the mean-field dynamics. The latter is obtained by integrating the deterministic equations (\ref{eq:barrio1},\ref{eq:barrio2}) using an Euler-forward scheme ($\Delta t=0.1$).  Model parameters are $P_0=100, \alpha_m=\alpha_p=1, \mu_m=\mu_p=0.03$. The figure is to be compared with Fig. 8 of \cite{barrio}.}
\label{fig:fig5}
\end{figure}

The resulting power spectra for a given set of parameters used e.g. in
\cite{barrio} are shown in Fig. \ref{fig:fig4}, and as seen in the
figure direct simulations based on a modified Gillespie algorithm,
similar to the one described in the section of the toy model, agree
well with the theoretical predictions. Remaining discrepancies are presumably due to finite-size and equilibration effects. We note that the peak of the
spectra shown in Fig. \ref{fig:fig4} occurs at an angular frequency of
roughly $\omega\approx 0.05$ in units of $1/$min , corresponding to a
time period of $T\approx 125$min, i.e. approximately two hours and
therefore close to the results from experiments reported e.g. by
Hirata et al. \cite{hirata}. This agreement is of course a consequence
of the specific choice of parameters, but it demonstrates that the
oscillatory behaviour of mRNA and protein concentrations during somite
segmentation may well be described as an effect of coherently
amplified intrinsic noise, as opposed to cycles produced by a
deterministic model. This enlarges the range of permissible model
parameters, and our analysis as well as that of \cite{barrio,bratsun} may
therefore provide additional support for the applicability of this
simple stochastic model. Our theoretical analysis may also be used in
order to test the robustness of the model without performing costly
stochastic simulations throughout a large range of parameter values. In
Fig. \ref{fig:fig5} for example we depict the frequency at which the
spectrum of mRNA fluctuations has its maximum in dependence on the
time delay $\tau$. This data is to be compared with Fig. 8 of
\cite{barrio}, where quantitatively very similar results were obtained
from actual stochastic simulations. Care needs to be taken though in
interpreting the maximum of the power spectra as the frequency at
which the system oscillates. The peaks in the spectra can be broad and
hence several modes contribute. Also, of course, finite systems at
small sizes may show deviations from the curves obtained from the system-size expansion, as the latter curves, even though they represent the next-to-leading order in $\Omega^{-1/2}$, can only be expected to be accurate at large system sizes. Still, Fig. \ref{fig:fig5} provides a
theoretical confirmation of the findings of \cite{barrio}, and
suggests that $2$-hour cycles are found at values of the delay time of about $\tau\approx 5-10$min
at $h=3$ and at slightly larger values of $\tau\approx 15-17$min at
$h=4$. It should be noted however that the variation of the observed
time period is rather small as $\tau$ and $h$ are varied in
Fig. \ref{fig:fig5} so that other parameter ranges are not ruled out
by the experimentally observed $2$hr period. Still, with theoretical approaches available along the lines discussed in the present paper or along those of \cite{bratsun} the most efficient way of identifying parameter values compatible with measurements in real-world experiments might be to use analytical expressions of the type presented in Eqs. (\ref{eq:spec1},\ref{eq:spec2}) first to narrow down the range of possible parameter values. Such closed form expressions can be evaluated relatively quickly and this pre-selection of model parameters based on analytical results is hence much less costly than performing parameter scans in stochastic simulations. Once suitable parameters have been identified from the theory, subsequent stochastic simulations in a much smaller
range of parameters can then be carried out to confirm whether or not the experimentally observed
behaviour is indeed found in the stochastic system.

%%%%%%%%%%%%%%%%%%%%%%%%%%%%%%%%%%%%%%%%%%%%%%%
\section{Discussion and concluding remarks}
In summary we have successfully extended recent analyses of the
effects of intrinsic noise to chemical systems with delay. In
particular we have shown that the picture of coherent oscillations,
induced by the discreteness of the microscopic
dynamics applies to systems with delay as well. Stochastic
self-sustained oscillations can here be found in {\em finite} systems
at choices of the model parameters for which the {\em infinite}
system, described by deterministic mean-field equations, does not
exhibit cycles. This observation has important implications for the
modelling of oscillatory biological systems with delay, as reaction
rates are often not known experimentally, but are instead tuned in
theoretical approaches, in order to ensure that simple model systems
reproduce the experimentally observed oscillatory behaviour. Our
analysis shows that confining model parameters to permissible ranges
in which the deterministic model shows oscillations may be
unnecessarily restrictive, as the stochastic dynamics at finite
system size may well exhibit oscillatory behaviour outside these
ranges of the model parameters.

While this observation as such has been made previously e.g. in
\cite{barrio,bratsun} the main contribution of the present work is the
extension of van Kampen expansion techniques to the case of stochastic
delay systems, and based on the resulting Langevin equation the
analytical calculation of the power spectra of fluctuations about
mean-field fixed points in delay systems. To our knowledge this has
not been attempted before, even though previous work on Kramers-Moyal
expansions in delay systems can be found in \cite{frank2}. Our
approach is here complementary to that of \cite{bratsun}, who have
used generating function techniques to study master equations of delay
systems, and to computer power spectra in good agreement with simulations, but who, in our understanding,
have not carried out a systematic expansion in the inverse system
size. Since the two approaches each rely on a series of approximations and simplifications analytical expressions derived in the two formalisms may not be fully equivalent. The spectra derived in our work are however in excellent agreement with
simulations, confirming the validity of the procedure carried out here. We have here first developed the general theory in the
context of a simple one-dimensional system. Generalisation to more
complex models with a higher number of degrees of freedom is possible
however, and as a further example we have addressed a basic model of
gene regulation. In particular we have studied a delay-system
describing the regulatory processes underlying the expression of $hes
1$ mRNA and Hes 1 protein. Simulational work has here recently been
reported by Barrio et al. \cite{barrio}, and our work complements
these mostly computational studies by an analytical computation of the
spectra of the observed oscillations in mRNA and protein expression
levels. See again also \cite{bratsun} for related models. Based on our analytical results
further characterisation of the behaviour of the model is possible,
without the need to perform computationally expensive simulations in a
wide range of parameters. Our theoretical approach is furthermore
applicable more generally, and can be expected to be useful for the
theoretical understanding of the behaviour more intricate stochastic delay systems.

%%%%%%%%%%%%%%%%%%%%%%%%%%%%%%%%%%%%%%%%%%%%%%%%%%%%%%%%%%%%%%%%%%

\begin{acknowledgments} 
TG is an RCUK Fellow (RCUK reference EP/E500048/1), and would like to thank R. Schlicht for useful discussions on stochastic delay systems, and A.~J. McKane and R.~P. Boland for earlier collaboration on systems with intrinsic fluctuations.
\end{acknowledgments}


\begin{thebibliography}{10}
\bibitem{discl} The numerical integration of the deterministic equation is performed with a simple Euler-forward scheme. The initial oscillation amplitude of the mean-field trajectory may here depend slightly on the chosen time-stepping. We have typically used $\Delta t=0.1$ or smaller. The asymptotic fixed point is not affected.
\end{thebibliography}
\end{document}